# Modification of Graphene Properties due to Electron-Beam Irradiation

D. Teweldebrhan and A.A. Balandin[*]

Nano-Device Laboratory, Department of Electrical Engineering, University of California – Riverside, Riverside, California 92521 USA

Materials Science and Engineering Program, Bourns College of Engineering, University of California – Riverside, Riverside, California 92521 USA

**Abstract**

The authors report micro-Raman investigation of changes in the single and bi-layer graphene crystal lattice induced by the low and medium energy electron-beam irradiation (5 – 20 keV). It was found that the radiation exposures results in appearance of the strong disorder $D$ band around 1345 cm$^{-1}$ indicating damage to the lattice. The $D$ and $G$ peak evolution with the increasing radiation dose follows the amorphization trajectory, which suggests graphene's transformation to the nanocrystalline, and then to amorphous form. The results have important implications for graphene characterization and device fabrication, which rely on the electron microscopy and focused ion beam processing.

---

[*] Corresponding author; E-mail address: balandin@ee.ucr.edu ; web-address: http://www.ndl.ee.ucr.edu

Graphene, a planar single sheet of $sp^2$-bonded carbon atoms arranged in honeycomb lattice, has attracted major attention of the research communities owing to a number of its unique properties [1-5]. From the practical point of view, some of the most interesting properties are the extraordinarily high room temperature (RT) carrier mobility up to ~27000 $cm^2V^{-1}s^{-1}$ [1, 3], and the recently discovered extremely high thermal conductivity exceeding ~3080 W/mK [5-6]. The outstanding current and heat conduction properties are beneficial for the proposed electronic, interconnect, and thermal management applications of graphene.

Graphene characterization and device fabrication often require an extensive use of the scanning electron microscopy (SEM), transmission electron microscopy (TEM), and focused ion beam (FIB) processing. These techniques involve electron beam (e-beam) irradiation of the samples, which may result in damage and disordering. The radiation induced defects may lead to significant deterioration of the electron and heat conduction properties. The damage to the material, which consists of a single or few atomic layers, can be quite significant even at low radiation doses. Despite the practical importance of this issue, no investigation of the e-beam effects on graphene properties has been reported so far. In addition to the practical significance, the influence of the irradiation on graphene properties is of fundamental science interest. A variety of physical properties of carbon nanostructures can be obtained by influencing their lattice structure. One of the methods for modifying the properties of carbon materials is the formation of defects through e-beam irradiation [7]. The effect of the irradiation on bulk graphite has been studied extensively due to graphite applications in thermal nuclear reactors [8-9]. At the same time, no irradiation data is available for single-layer graphene (SLG) or bi-layer graphene (BLG).

In this letter we report investigation of the modification in graphene induced by the low and medium energy e-beams. The possible changes due to irradiation include the transformation of the crystalline lattice into nanocrystalline or amorphous; displacement of atoms from the lattice; excitation of phonons and plasmons, which results in the sample heating. Any of these irradiation effects will lead to modification of the phonon modes. For this reason, we selected the micro-Raman spectroscopy as a main characterization tool. Raman spectroscopy methods are capable for detecting small changes in the crystal structure and have been used extensively in the analysis of the irradiation damage on other carbon materials [10-11].

Graphene samples were produced by the mechanical exfoliation from three different types of initial carbon material: (i) bulk highly oriented pyrolitic graphite (HOPG), (ii) Kish graphite and (iii) high-pressure high-temperature (HPHT) graphitic layers. Graphene HPHT synthesis was recently reported by us [12]. We did not observe a systematic difference in the properties of graphene produced by these different techniques. All graphene flakes were placed on the standard Si/SiO$_2$ substrates and initially identified with an optical microscope. SLG and BLG sample were selected using micro-Raman spectroscopy through the *2D*-band deconvolution [13-15]. The samples were cleaned using the standard procedure and kept under vacuum conditions (~$10^{-4}$ Torr or below) to reduce the organic and vapor contamination. A typical sample is shown on SEM image in Fig 1. Raman spectra were measured using a Renishaw spectrometer under 488-nm laser excitation in the backscattering configuration. In order to prevent local heating the excitation power was kept below 2.0 mW [16].

The low-energy (5 keV) and medium-energy (20 keV) electron irradiation was performed with the help of the Philips XL-30 FEG field-emission system. The graphene samples were subjected to continuous e-beams from the electron gun focused on an area of $1.6 \times 10^9$ nm$^2$ in vacuum. Under the beam current of ~0.15 nA, measured with the Faraday cup, the dose density rate is ~0.59 e$^-$/nm$^2$s for the 20-keV beam. A constant emission current of 235 μA was maintained during the exposure. The working distance between the samples and the tip of the electron gun was kept at ~6.0 mm. The flux has been maintained constant for each of the experiments so that the electron dose was proportional to the irradiation time. The dose density for SLG flake after 30 min irradiation was ~$1.06 \times 10^{17}$ e$^-$/cm$^2$. The areal density of carbon atoms is ~$7.6 \times 10^{-8}$ g/cm$^2$.

Fig. 2 shows the evolution of Raman spectra of SLG under e-beam irradiation. These spectra manifest two important features: the disorder *D* band and zone-center *G* peaks near ~1345 cm$^{-1}$ and ~1580 cm$^{-1}$, respectively. In pristine graphene the disorder *D* peak is absent. The latter indicates the high quality of graphene and its crystalline nature. The *D* peak is attributed to A$_{1g}$ symmetry phonons near the *K*-zone boundary. These phonons are not Raman active due to the momentum conservation in the scattering. They became active in the presence of structural disorder as described by the double – resonance model [17-18]. The *D* peak appears after very short exposure time in the spectra of both SLG and BLG. Irradiation by the low and medium energy e-beams leads to similar results. The intensity of the *D* peak initially grows, attains its maximum within a few minutes of the e-beam exposure and then decreases with the increasing dose of irradiation. The full width at half maximum (FWHM) for *D* and *G* broadens under irradiation. At short exposure time one can notice an appearance of another disorder related peak at ~1620 cm$^{-1}$. This *D'* peak has been observed in defected graphite [19]. In Fig. 3 we present the second-order Raman spectrum of pristine and irradiated SLG. The main features are the *2D* band at ~2685 cm$^{-1}$ and *2D'* band around 3240 cm$^{-1}$. The broad band around 2930 cm$^{-1}$ is attributed to *D+G* overtone. The evolution of the second-order spectrum of BLG under irradiation is similar to that of SLG at both 5 and 20 keV. The changes in all graphene spectra under irradiation are indicative of the disorder and defects introduced due to the e-beam.

In order to rationalize the results we plotted the ratio of *D* and *G* peak intensities *I(D)/I(G)* as a function of the e-beam irradiation time (see Fig. 4). After the first few minutes of irradiation the ratio *I(D)/I(G)* attains its maximum and then falls rapidly. The continuation of the irradiation results in a slower decrease or saturation of *I(D)/I(G)*. This trend was observed for both SLG and BLG. It is illustrative to compare this plot with the amorphization trajectory proposed by Ferrari and Robertson for carbon materials [20-21]. They considered that the Raman spectrum of all carbons depend on (i) clustering of the sp$^2$ phase; (ii) bond disorder; (iii) presence of sp$^2$ rings/chains; and (iv) sp$^2$/sp$^3$ ratio. The trends summarized by the amorphization trajectory indicate that *I(D)/I(G)* increases when crystalline graphite evolves into nanocrystalline (*nc*) graphite (stage I); and then decreases when *nc* graphite becomes mainly-sp$^2$ amorphous carbon (stage II). Our case appears to follow the first two stages of the amorphization trajectory, which suggest that crystalline graphene under irradiation transforms into *nc* phase, possibly with localized defects, and then, as the radiation damage increases, becomes more disordered, i.e. amorphous.

The stage III in the original amorphization trajectory for bulk graphite [20] is characterized by farther decrease in *I(D)/I(G)*, which corresponds to the increase in sp$^3$ content and formation of the tetrahedral amorphous carbon. The situation is different for irradiated graphene where *I(D)/I(G)* tends to saturate with increasing radiation dose (see Fig. 4). The latter can be related to the fact that we deal with just one or two atomic layers of materials and $sp^3$ phase does not form easily. The threshold acceleration voltage of knock-on damage, i.e. ballistic ejection of an atom, for single-wall carbon nanotubes (CNTs) is about 86 keV [22]. Assuming that the threshold is similar for graphene and taking into account that in our experiments the electron energy was below 20 keV, we can exclude the vacancies due to the knock-on damage as possible mechanism for the observed lattice modification. Formation of mechanical cracks was not responsible for the spectrum modification. It was verified by measuring spectra from many spots on the sample and extending exposure time to several hours. It was found that mechanical cracks only appear after 2 hours of e-beam exposure. Possible carbon residue on the surface due to disassociation of carbon- containing molecules is not expected to strongly affect our results because of the cleaning treatment and vacuum conditions of the experiment. The molecular residue signatures can also be distinguished from the regular *G* peak [23].

As an independent confirmation of the graphene lattice modification we measured current – voltage characteristics of graphene before and after irradiation. The measurements were performed for the standard top-electrode back-gated structures. It was found that the resistance drastically increases as graphene samples go through stages I and II (see inset to Fig. 4) suggesting an evolution to the *nc* and amorphous forms. We can estimate the apparent *nc* size in graphene lattices after few minutes of e-beam irradiation using Tuinstra – Koening [24] relation $I(D)/I(G)=C(\lambda)/L_a$, where $C(\lambda=488nm) \sim 4.4$ nm [21, 24-25] and $L_a$ is the cluster size or in-plane correlation length. In our case, at the end of stage I, the expected grain size is on the order of $L_a \sim 2.4 - 3.5$ nm. The increase in the irradiation dose results in conversion into mainly-$sp^2$ amorphous carbon. One may consider it to be unexpected that rather a short-time exposure of graphene to the low or medium-energy e-beams results in such damage. The reported Raman spectroscopy measurements indicate substantial damage to the single-wall CNTs for the low-energy e-beam dose of $\sim 8 \times 10^{17}$ e/cm$^2$ [26]. Similar effects were observed for CNTs even with lower energies and beam currents in the presence of water vapour [27]. The irradiation dose required for damaging graphene is smaller than that for CNTs possibly because of graphene's flat geometry, which makes it more susceptible to the electron flux. Our results have important consequences for graphene device fabrication where SEM imaging is involved. Graphene, a perfect conductor of electricity and heat, can be converted to the electrical and thermal insulators [28] by e-beam irradiation during the SEM and FIB fabrication steps.

*Acknowledgements*

This work was supported, in part, by SRC - DARPA Focus Center Research Program (FCRP) through Functional Engineered Nano Architectonics (FENA) Center and Interconnect Focus Center (IFC). The authors acknowledge the help of Dr. K.N. Bozhilov and Mr. G. Liu with the measurements and thank Dr. K.S. Novoselov for his valuable comments on the manuscript.

**Figure Captions**

Figure 1: Scanning electron microscopy image of graphene flake with the single and bi-layer regions.

Figure 2: Raman spectrum of SLG under electronic beam irradiation.

Figure 3: Second-order Raman spectrum of SLG under electronic beam irradiation.

Figure 4: Evolution of the ratio of the intensities of the D and G peaks as a function of the irradiation exposure for SLG and BLG.

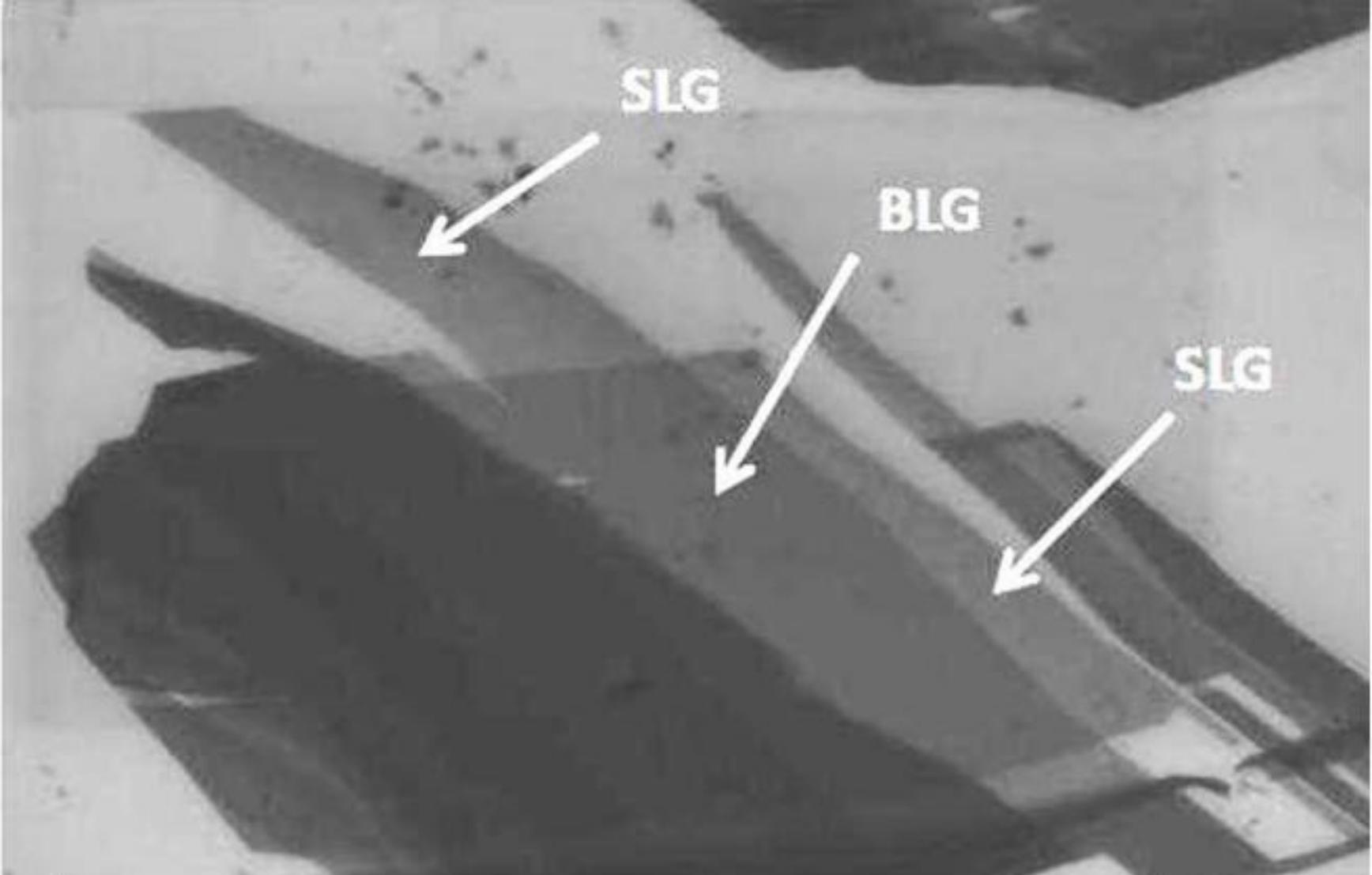

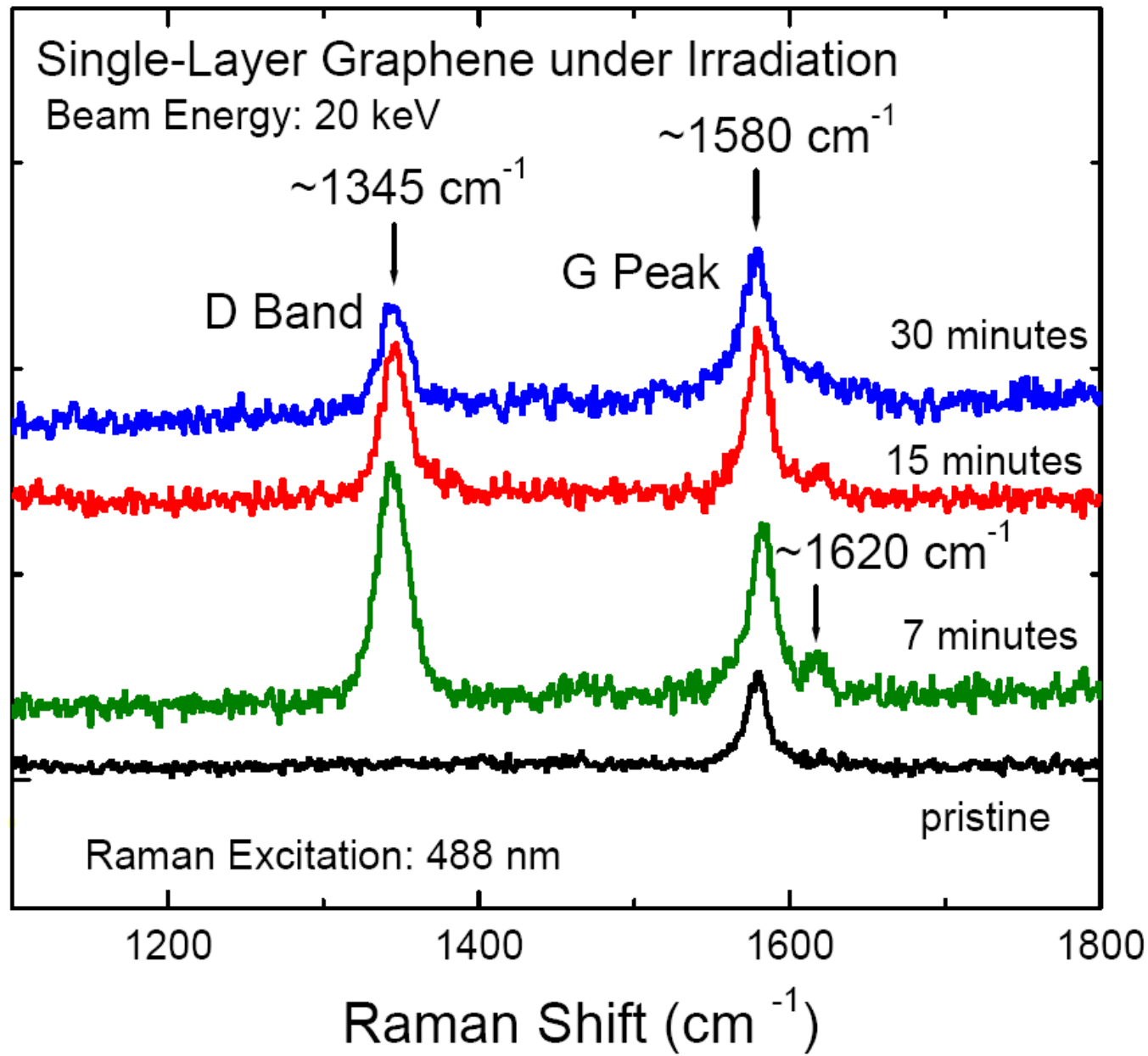

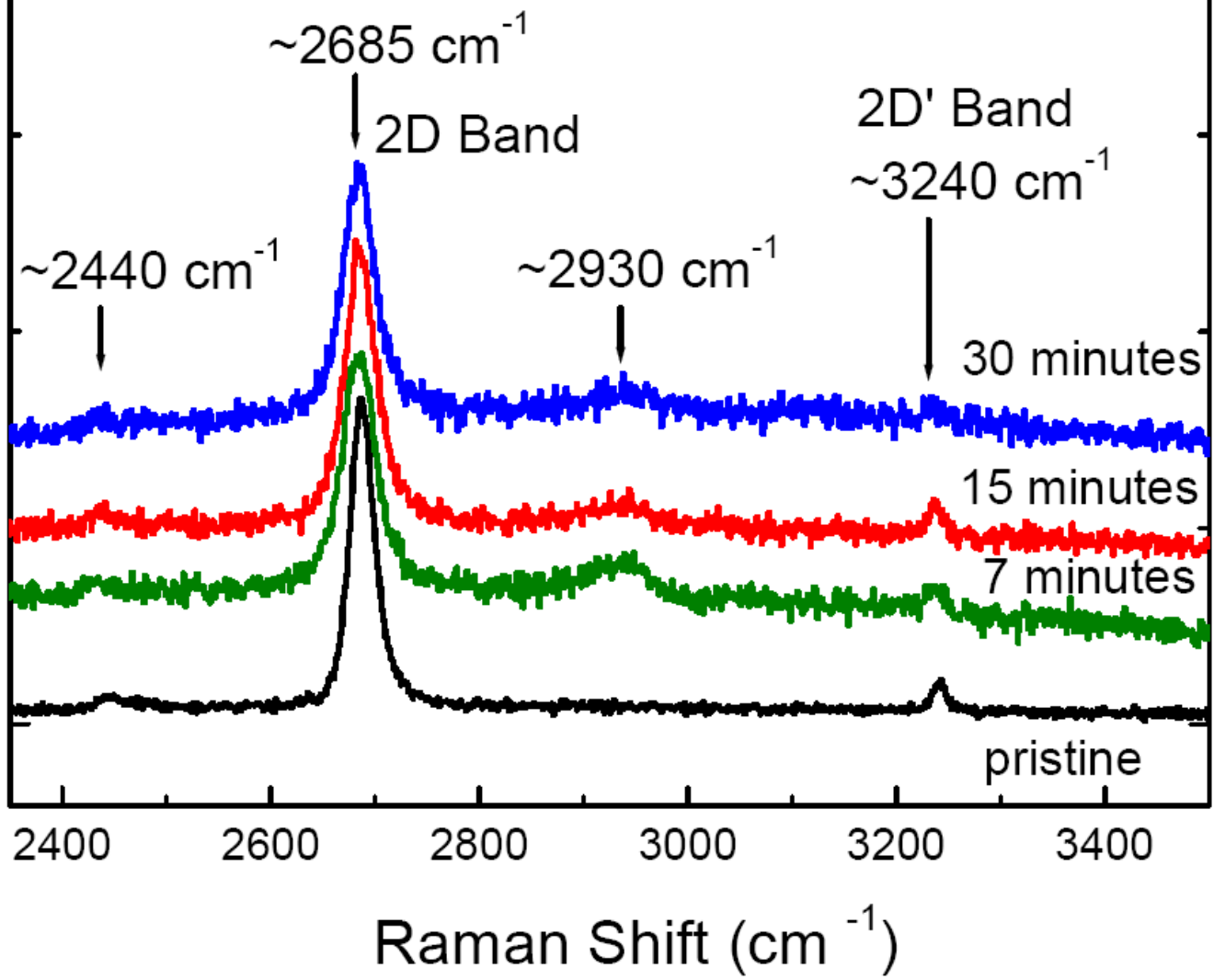

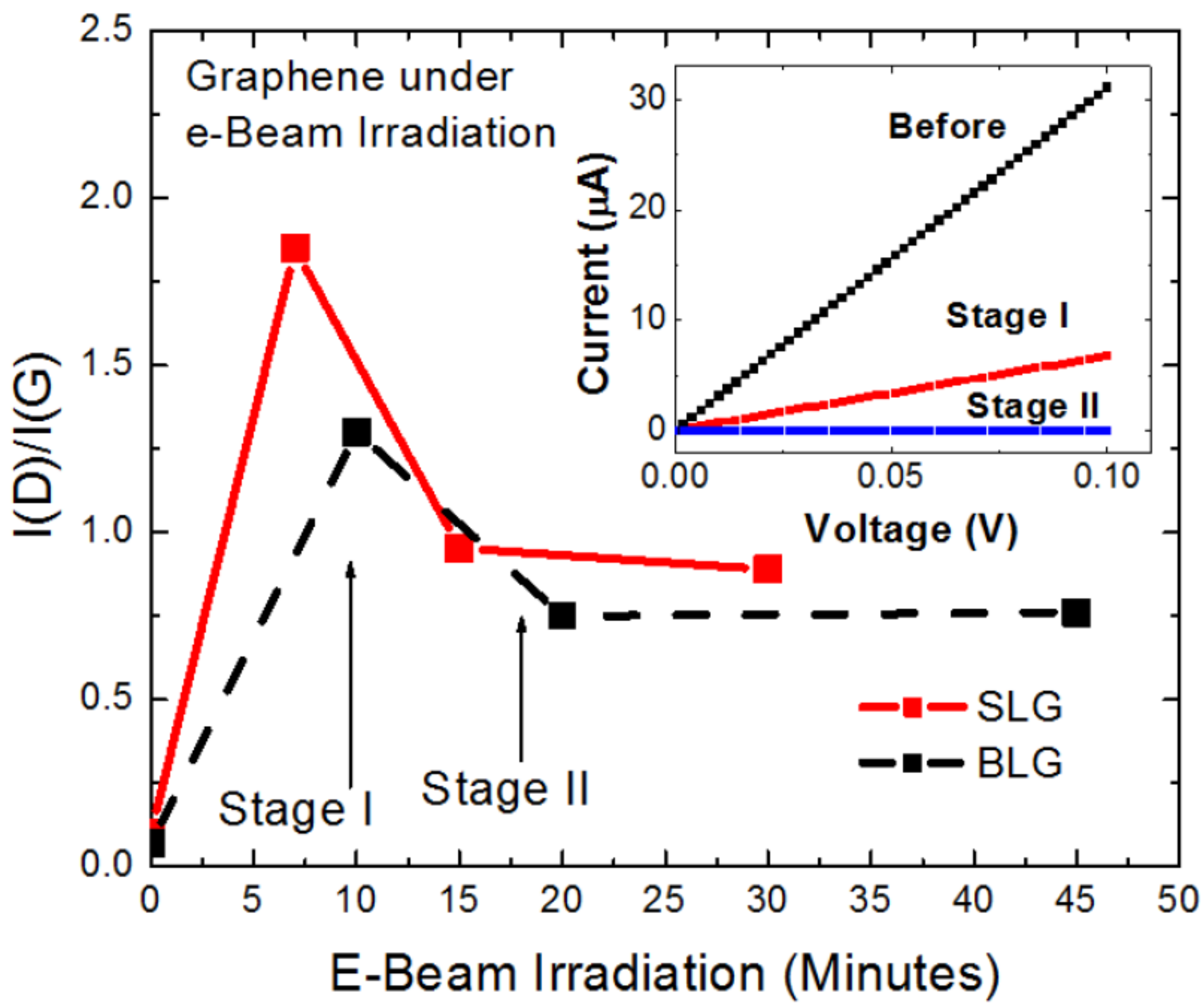